\documentclass[draft,aps,twocolumn,prl,superscriptaddress,nofootinbib]{revtex4}
\usepackage{amsmath}
\usepackage{amsfonts}
\usepackage[centertableaux,boxsize=0.35cm]{ytableau}

\DeclareMathOperator{\ch}{ch}
\DeclareMathOperator{\Conf}{Conf}
\DeclareMathOperator{\Gr}{G}
\DeclareMathOperator{\Newt}{Newt}
\DeclareMathOperator{\Trop}{Trop}

\begin{document}

\title{Non-perturbative geometries for planar $\mathcal{N}=4$ SYM amplitudes}

\author{Nima Arkani-Hamed}
\affiliation{School of Natural Sciences, Institute for Advanced Study, Princeton, NJ 08540}
\affiliation{Center of Mathematical Sciences and Applications, Harvard University, Cambridge, MA 02138}
\author{Thomas Lam}
\affiliation{Department of Mathematics, University of Michigan, Ann Arbor, MI 48109}
\affiliation{Department of Mathematics, Massachusetts Institute of Technology, Cambridge, MA 02139}
\author{Marcus Spradlin}
\affiliation{Department of Physics and Brown Theoretical Physics Center, Brown University, Providence, RI 02912}

\begin{abstract}
There is a remarkable well-known connection between the $\Gr(4,n)$ cluster algebra and $n$-particle amplitudes in $\mathcal{N}=4$ SYM theory. For $n \ge 8$ two long-standing open questions have been to find a mathematically natural way to identify a finite list of amplitude symbol letters from among the infinitely many cluster variables, and to find an explanation for certain algebraic functions, such as the square roots of four-mass-box type, that are expected to appear in symbols but are not cluster variables. In this letter we use the notion of ``stringy canonical forms'' to construct poly\-topal realizations of certain compactifications of (the positive part of) the configuration space $\Conf_n(\mathbb{P}^{k-1}) \cong$ $\Gr(k,n)/T$ that are manifestly finite for all $k$ and $n$. Some facets of these polytopes are naturally associated to cluster variables, while others are naturally associated to algebraic functions constructed from Lusztig's canonical basis. For $(k,n) = (4,8)$ the latter include precisely the expected square roots, revealing them to be related to certain ``overpositive" functions of the kinematical invariants.
\end{abstract}

\maketitle

\section{I. Introduction}

Scattering amplitudes are boundary observables in flat space, depending only on the kinematical data specifying the helicities and momenta of the scattering particles. It is thus natural to ask whether there is some question that can be posed in kinematic space whose answer yields the amplitudes directly, without referring to auxiliary notions such as unitary evolution in the bulk of spacetime or a string worldsheet. The challenge appears daunting since there are no obvious physical notions of locality or time associated with, say, the space of $n$ null momenta relevant for the scattering of massless particles. We must instead cast out more adventurously, looking for new sorts of mathematical structures in this naively barren space, with the power to generate all of the richness and complexity needed for scattering amplitudes compatible with locality and unitarity. The past several years have seen significant inroads in this program, associated with deep new combinatorial and geometric structures in kinematic space connecting various aspects of amplitudes to mathematical notions of total positivity, cluster algebras and motives in startling new ways.

A prototype of the kind of description we seek is provided for the all-loop integrand in planar $\mathcal{N}=4$ super-Yang-Mills (pSYM) theory by the amplituhedron~\cite{Arkani-Hamed:2013jha}, which can be understood purely in the kinematical momentum-twistor space: the (super)integrand is the unique canonical form~\cite{Arkani-Hamed:2017tmz} with logarithmic singularities on (and only on) all boundaries of the amplituhedron. Thus the integrand is fully determined by some geometry in kinematic space (the amplituhedron) and a question asked of that geometry (the determination of its canonical form). The amplituhedron provides a geometric origin for all of the singularities of the integrand as a rational function. But the simplicity gained in dealing with rational functions comes at a significant cost: the amplituhedron is inexorably tied to perturbation theory. Indeed, there is a different geometry for every loop order.

For the full amplitude we should instead expect some geometric origin for the much more intricate pattern of branch cuts, which are present non-perturbatively. Of course the question of determining the geometry of branch cuts from first principles, for instance from an analysis of Landau equations, was an infamously difficult one in the S-matrix program in the 1960s. But there is hope for planar theories of massless particles, like pSYM theory, where it is known that for any fixed particle number, the number of branch points associated with solutions to the Landau equations is finite~\cite{Prlina:2018ukf}. We also expect that perturbation theory for the planar theory has a finite radius of convergence.

Given this encouragement, there is a natural candidate for the non-perturbative geometry we seek. The kinematic data is provided by $n$ momentum-twistor four-vectors~\cite{Hodges:2009hk} $Z_1^I, \ldots, Z_n^I$, and the action of the conformal group SL(4) tells us we can associate this with a point in the Grassmannian $\Gr(4,n)$, a $4(n{-}4)$-dimensional space. Restricting for simplicity to the case of MHV amplitudes, the amplituhedron asks for the external data to lie in the {\it positive} Grassmannian $\Gr_+(4,n)$~\cite{ArkaniHamed:2012nw}, so the positivity of external data should clearly be an important ingredient. Even more concretely, there is apparently a fascinating connection between the positivity of kinematic data and the Landau equations---in all examples studied to date, the Landau equations admit {\it no} solutions when the external data is taken to be in $\Gr_+(4,n)$! This is a highly nontrivial fact, implying that amplitudes have no branch points inside the positive domain, again suggesting that this region should play a starring role in defining the non-perturbative geometry relevant to amplitudes.

MHV amplitudes enjoy an additional little group symmetry under which $Z_i \to t_i Z_i$, and as is familiar, they are therefore functions of cross-ratios, depending only on $3(n{-}5)$ rather than $4(n{-}4)$ variables. Thus an obvious guess for the non-perturbative geometry is ``$\Gr_+(4,n)$ modulo the the little group torus", denoted $\Gr_+(4,n)/T$.

Indeed, quite apart from this more recent motivation involving positivity, it has long been appreciated~\cite{Golden:2013xva} that there is a deep connection between MHV amplitudes and the configuration space $\Conf_n(\mathbb{P}^3) \cong \Gr(4,n)/T$. All evidence available to date from explicit multi-loop computations~\cite{Goncharov:2010jf,Caron-Huot:2019bsq,Dixon:2016nkn} supports the hypothesis that the symbol alphabets of $n=6,7$ particle amplitudes are the cluster variables of the $\Gr(4,n)$ Grassmannian cluster algebra. This has long been a source of inspiration for determining the non-perturbative geometry of pSYM theory, but for $n \ge 8$ there are a couple of open questions indicating that the cluster algebra itself is not the end of the story.

First, as we have stressed, we expect that the number of branch points for all pSYM amplitudes, and hence the number of symbol letters, should be finite for any $n$. However, the $\Gr(k,n)$ cluster algebra has infinitely many cluster variables if $(k{-}2)(n{-}k{-}2) > 3$, including the cases $k = 4$, $n \ge 8$ of interest to amplitudes. Is there a mathematically natural way to extract some ``finite subset'' of variables relevant to scattering amplitudes?

Second, cluster variables for $\Gr(k,n)$ are always polynomials in minors of the $Z$ matrix. But beginning with $n=8$, even for MHV amplitudes that should be polylogarithmic, we expect (specifically, starting at three loops~\cite{Prlina:2017azl,Prlina:2017tvx}) symbol letters that are algebraic functions of these minors. The most familiar and famous of these is the square root associated with the four-mass box integral~\cite{Hodges1977,tHooft:1978jhc}, but a variety of algebraic letters appear in various contexts (see for example~\cite{Chicherin:2017dob,Gehrmann:2018yef,Chicherin:2018yne,Heller:2019gkq,Zhang:2019vnm}, and~\cite{Besier:2019kco,Bourjaily:2019igt} for some tools for dealing with them). How can we see them arise in a mathematically natural way?

One might think that defining $\Gr_+(k,n)/T$ is completely straightforward, and indeed there is no subtlety associated with thinking of what the {\it interior} of this space. But in the S-matrix program it is precisely the boundaries of kinematic space that are of particular interest, since these are where amplitudes can have singularities. So from the perspective of determining a non-perturbative geometry for amplitudes, it is imperative to understand the {\it boundary} structure of $\Gr_+(k,n)/T$, and this involves making a choice of compactification. In this letter we propose natural compactifications of $\Gr_+(k,n)/T$ that address both of the above questions, providing us (when $k=4$) with candidate non-perturbative geometries for the amplitudes of pSYM theory. We provide explicit polytopal realizations of these compactifications using the ``stringy canonical forms" of~\cite{AHL}. This construction has a number of connections to other ideas, such hypersimplex decompositions and tropical Grassmannians, and we defer a systematic exposition to a longer companion paper~\cite{toappear}. Our purpose in this letter is instead to summarize some of the essential ideas and results relevant to pSYM theory.

In Sec.~II we introduce certain compactifications of $\Gr_+(k,n)/T$ that manifestly have a finite number of facets for any $k, n$. In cases when the corresponding cluster algebra is finite each facet is naturally associated with a cluster variable, but in infinite cases there are additional facets. In Sec.~III we describe a natural way of associating cluster algebraic functions, that generalize the notion of cluster variables, to such facets. In Sec.~IV we study the case $(k,n) = (4,8)$ in detail. We find that the ``extra'' facets are associated with the famous square root associated to the four-mass box. Ancillary files contain data pertaining to several of the polytopes we study.

\section{II. Kinematic Space Polytopes}

Scattering amplitudes of $n$ particles in pSYM theory are functions on $\Conf_n(\mathbb{P}^3)$, the configuration space of $n$ points in $\mathbb{P}^3$. (Strictly speaking this is true only for MHV amplitudes; non-MHV amplitudes are most naturally thought of as differential forms~\cite{Arkani-Hamed:2017vfh} on a $(\mathbb{C}^*)^{n-1}$ bundle over $\Conf_n(\mathbb{P}^3)$.) There is a birational isomorphism~\cite{Golden:2013xva} $\Conf_n(\mathbb{P}^{k-1}) \simeq \Gr(k,n)/(\mathbb{C}^*)^{n-1}$ because generic points in this configuration space can be represented by maximal rank $k \times n$ matrices $Z$, with the columns representing the homogeneous coordinates of $n$ points in $\mathbb{P}^{k-1}$, modulo SL($k$) and modulo independent rescaling of each column.

The (open) \emph{positive domain} of $\Conf_n(\mathbb{P}^{k-1})$ (with respect to the ordering $1,\ldots,n$) is defined as the positive Grassmannian $\Gr_+(k,n)$ modulo the torus action $T = \mathbb{R}_+^n$ that rescales columns. The problem before us is that of understanding the boundary structure of this domain under a suitable compactification. There exist many inequivalent compactifications, with the choice appropriate for any particular application determined by, indeed one should say defined by, the class of functions under consideration. Ultimately it is the scattering amplitudes themselves that dictate the compactification of $\Gr_+(4,n)/T$ that is relevant to pSYM theory.

The case $k=2$ is well-known to both mathematicians and physicists: $\Gr_+(2,n)/T$ is the moduli space of $n$ ordered points on the real line, and its Deligne-Mumford compactification~\cite{DM} has long been known~\cite{Koba:1969kh} to underlie the structure of open string amplitudes. The most natural generalization to $k>2$ is the positive Chow quotient of the Grassmannian~\cite{toappear}, which we will refer to as the \emph{totally nonnegative configuration space} $\overline{\Gr_+(k,n)/T}$. Geometrically, it is the closure of $\Gr_+(k,n)/T$ inside the Chow quotient of the Grassmannian $\Gr(k,n)/\!/T$~\cite{Kapranov}. In~\cite{AHL} it has recently been shown that Koba-Nielsen-like string worldsheet integrals can be used to construct polytopal realizations of various positive spaces, including $\overline{\Gr_+(k,n)/T}$, generalizing the well-known realization of the compactification of $\Gr_+(2,n)/T$ as the $A_{n-3}$ associahedron. The space $\overline{\Gr_+(k,n)/T}$ comes with a stratification that will be discussed elsewhere~\cite{toappear}, with strata labeled by several combinatorially interesting data including positroid decompositions of the $k,n$ hypersimplex, faces of the tropical positive Grassmannian, positive tropical Pl\"ucker vectors, etc. In this note we content ourselves with presenting the simplest data about its polytopal realization, which we denote by $\mathcal{C}(k,n)$.

Here we briefly review the key steps of~\cite{AHL}. A \emph{positive parameterization} is a $d=(k{-}1)(n{-}k{-}1)$-parameter family of $k \times n$ matrices $Z(x_1,\ldots,x_d)$ that covers all of $\Gr_+(k,n)/T$ as the parameters range over the positive orthant $(x_1,\ldots,x_d) \in \mathbb{R}^d_+$. In the following it will be important that we always use a \emph{cluster parameterization}, which means that the parameters are Fock-Goncharov coordinates in the initial cluster of the $\Gr(k,n)$ cluster algebra (for which we use the conventions of Fig.~2 of~\cite{CDFL}). For example,
\begin{align}
Z = \begin{pmatrix}
0 & 1 & 1 & 1 & 1 \\
-1 & 0 & 1 & 1 + x_1 & 1 + x_1 + x_1 x_2
\end{pmatrix}
\end{align}
is a cluster parameterization of $\Gr_+(2,5)/T$ because
\begin{align}
\frac{\langle 12 \rangle \langle 34 \rangle}{
\langle 14 \rangle \langle 23 \rangle} = x_1\,, \qquad
\frac{\langle 13 \rangle \langle 45 \rangle}{
\langle 15 \rangle \langle 34 \rangle} = x_2\,.
\end{align}
We use $\langle i_1 i_2 \ldots i_k \rangle = \det(Z_{i_1} \cdots Z_{i_k})$ to denote Pl\"ucker coordinates on $\Gr(k,n)$, where $Z_i$ is the $i$th column of $Z$. The canonical form~\cite{Arkani-Hamed:2017tmz} on $\Gr_+(k,n)/T$ is $\Omega = \bigwedge d\log x_i$.

If $P$ is a homogeneous polynomial in Pl\"ucker coordinates we let $\Newt(P)$ denote the Newton polytope of $P$ in $\mathbb{R}^d$ respect to the variables $(x_1,\ldots,x_d)$. The string integral construction associates to any sufficiently large (defined in~\cite{AHL}) collection $\mathcal{P} = \{P_1, \ldots, P_\ell\}$ of such polynomials a polytopal realization of a compactification of $\Gr_+(k,n)/T$ obtained by taking the convex hull of the Minkowski sum of the corresponding $\Newt(P_i)$.

We define $\mathcal{C}(k,n)$ to be the polytope obtained by taking $\mathcal{P}$ to be the set of all $\binom{n}{k}$ Pl\"ucker coordinates. If $k=2$ this gives the familiar realization~\cite{Arkani-Hamed:2017mur} of the $A_{n-3}$ associahedron. However for $k>2$, $\mathcal{C}(k,n)$ is different than the corresponding cluster polytope~\cite{FZY}, and is in particular a manifestly finite polytope for any $k$ and $n$, even when $(k{-}2)(n{-}k{-}2) > 3$ in which case the cluster algebra is infinite and it is not clear that there even exists a cluster polytope. Using~\cite{Minksum,Normaliz} we have computed the vertices, found the bounding hyperplanes, and analyzed the polyhedral combinatorics of $\mathcal{C}(k,n)$ for various $(k,n)$. Here for brevity we summarize just the $f$-vectors
\begin{align}
\begin{split}
(3,6):&\quad (1, 48, 98, 66, 16, 1)\,, \\
(4,7):&\quad (1, 693, 2163, 2583, 1463, 392, 42, 1)\,, \\
(3,8):&\quad (1, 13612, 57768, 100852, 93104, 48544, \\
& \qquad \qquad \quad \ \, \, \quad \qquad 14088, 2072, 120, 1)\,, \\
(4,8):&\quad (1, 90608, 444930, 922314, 1047200, 706042, \\
& \qquad \qquad\ \ \ \ \quad \quad 285948, 66740, 7984, 360, 1)\,.
\end{split}
\label{eq:gknmodt}
\end{align}
The first two of these have appeared as the duals of the fans associated to the tropical positive Grassmannian $\Trop^+ \Gr(k,n)$~\cite{SpeyerWilliams}, and some applications to physics for all four have recently been discussed in~\cite{Cachazo:2019apa,Drummond:2019qjk,Cachazo:2019ble}. Note that these polytopes are neither simple nor simplicial.

Except for the special case $n=6$, the $\mathcal{C}(4,n)$ polytopes are not invariant under the parity transformation $Z_i \mapsto *(Z_{i-1}Z_i Z_{i+1})$ that is a symmetry of MHV amplitudes. There are several ways of constructing parity-invariant polytopes, for example by taking a larger set $\mathcal{P}$ that includes the parity conjugates of all Pl\"ucker coordinates, or by taking a smaller set $\mathcal{P}$ of Pl\"ucker coordinates that is closed under parity. We find that the second option gives a particularly interesting polytope we call $\mathcal{C}^\dagger(k,n)$, obtained by taking $\mathcal{P}$ to be the subset of Pl\"ucker coordinates having the form $\langle i\,i{+}1\,j\,j{+}1\rangle$ or $\langle i\,j{-}1\,j\,j{+}1\rangle$. This is the largest subset of Pl\"ucker coordinates that is closed under parity, and for $n=7,8$ gives parity-invariant polytopes with $f$-vectors
\begin{align}
\begin{split}
(4,7):&\quad (1, 595, 1918, 2373, 1393, 385, 42, 1)\,,\\
(4,8):&\quad (1, 49000, 249306, 536960, 635176, 447284, \\
&\qquad\qquad\qquad\,\,\, \ 189564, 46312, 5782, 274, 1)\,.
\end{split}
\label{eq:unnamed}
\end{align}
Interestingly the number 595 appeared in~\cite{SpeyerWilliams}, where it was noted to be the number of facets of (the dual of) $\mathcal{C}(4,7)$ that are simplicial. However this seems to be a coincidence: we find that 50356 (not 49000) facets of (the dual of) $\mathcal{C}(4,8)$ are simplicial.

The virtue of using a cluster parameterization in the string integral construction is that it ties the geometry of the resulting polytope to the combinatorics of cluster algebras. Specifically, we find that the (outward) normal rays to all facets of $\mathcal{C}(k,n)$ are generated by ${\bf g}$-vectors~\cite{FZ4} of the $\Gr(k,n)$ cluster algebra in all of the cases listed above where the latter is finite. We remind the reader that the ${\bf g}$-vectors associated to the cluster variables in any one cluster generate a cone, the cones associated to different clusters are non-overlapping, and the union of all cones (called the ``cluster fan'') covers all of $\mathbb{R}^d$ in finite cases. One of the salient features of infinite cluster algebras is the last of these is no longer true: there are directions in $\mathbb{R}^d$ that are outside the cluster fan, and so are not associated to any cluster.

On the other hand the cones associated to the outward pointing normal rays to any polytope (the ``normal fan'') manifestly cover all of space. In infinite cases some normal rays to our polytopes point in directions outside the cluster fan, and it is interesting to identify their cluster-algebraic significance. Specifically, in the infinite case $(4,8)$ we find that the normal rays to 356 of the facets of $\mathcal{C}(4,8)$ lie along ${\bf g}$-vectors of the $\Gr(4,8)$ cluster algebra, consistent with the results reported in~\cite{Drummond:2019qjk,Southampton,HP} for the fan associated to $\Trop^+ \Gr(4,8)$. However, the remaining 4 normal rays are not even inside the cluster fan, according to the criterion given in Theorem 3.25 of~\cite{CDFL}. The normal rays to the 274 facets of $\mathcal{C}^\dagger(4,8)$ are a proper subset of those of the facets of $\mathcal{C}(4,8)$; 272 of them lie along ${\bf g}$-vectors and 2 of them are outside the cluster fan.

The cluster variables associated to the 272 ${\bf g}$-vector facets of $\mathcal{C}^\dagger(4,8)$ include the 108 symbol letters of the two-loop MHV amplitude~\cite{CaronHuot:2011ky} and the 64 additional rational letters of the two-loop NMHV amplitude~\cite{Zhang:2019vnm}. It would be interesting to see if this 272-letter alphabet exhausts the rational letters of (at least the MHV) 8-particle amplitudes to all loop order. (All evidence available to date suggests that the corresponding statement is true for the 9- and 42-letter symbol alphabets for the cases $n=6, 7$ respectively~\cite{Caron-Huot:2019bsq,Dixon:2016nkn}.)

In the following two sections we turn our attention to the remaining 2 (4) normal rays of $\mathcal{C}^\dagger(4,8)$ ($\mathcal{C}(4,8)$), which lie outside the cluster fan and so are not naturally associated to any cluster variables. We will see that the canonical basis element associated to the first integer point along each of these rays is overpositive, and we conjecture that the basis elements associated to points further along the rays are encapsulated in quadratic generating functions with positive roots. In the case of $\mathcal{C}^\dagger(4,8)$, these turn out to be precisely the roots associated to the four-mass box integral~\cite{Hodges1977,tHooft:1978jhc}.

\section{III. Cluster Canonical Bases}

Cluster variables in a rank $d$ cluster algebra are naturally associated with various lattice points in $\mathbb{Z}^d$, defined with respect to some initial seed cluster. One rather intuitive one is the notion of a ``denominator vector". The Laurent phenomenon tells us that every cluster variable can be expressed as a ratio polynomial/monomial in the initial seed cluster variables, and the denominator vector of a cluster variable is the exponent vector of the monomial appearing in this expression. A more canonical object is the ${\bf g}$-vector we have already alluded to; one can think of the {\bf g}-vector as the denominator vector with respect to an ordering defined by the $B$-matrix of the initial cluster. More precisely, one can define a partial ordering on the space of $n$-dimensional vectors {\bf g} by saying that ${\bf g}' \preceq {\bf g}$ iff ${\bf g}' -{\bf g}$ is in the cone spanned by the columns of the initial $B$-matrix. If we expand any cluster variable $x$ as a sum of Laurent monomials in initial cluster variables, the {\bf g}-vector of $x$ is defined to be that of the term whose {\bf g}-vector is smallest with respect to $\preceq$.

As we have mentioned, in finite-type cluster algebras, the {\bf g}-vectors associated to the cluster variables of any cluster define a cone, and these cones are remarkably non-overlapping and cover all of $\mathbb{R}^d$. This is related to another beautiful fact in finite type: the collection of monomials associated with all the clusters provides a basis for the entire cluster algebra. This is not obvious. The Laurent phenomenon guarantees that any product of cluster variables can be represented as a Laurent polynomial in terms of some initial seed cluster variables, but the claim is that any product of cluster variables from arbitrarily distant clusters can be written as a polynomial made of sums of monomials of variables in the same cluster.

Beyond finite type, while the cluster cones are still non-overlapping, they do not cover all of space. Related to this, cluster monomials no longer provide a basis for the full cluster algebra, as there are directions in {\bf g}-vector space that can not be spanned by products of cluster variables. There is a large literature on the construction of various ``canonical bases" of cluster algebras in infinite type. For $\Gr(k,n)$ every integer point in {\bf g}-vector space can be assigned~\cite{CDFL} (partly conjecturally) a polynomial in cluster variables, Lusztig's canonical basis element~\cite{Lusztig}. When the integer point lies inside a {\bf g}-vector cone of some cluster, the corresponding function is simply the obvious associated monomial in cluster variables, but when the integer point is on a ray not pointing in the direction of a cluster variable, the corresponding basis element is not a cluster variable, but something else.

We will encounter precisely this situation in our study of $\mathcal{C}(k,n)$, and explicitly for the $(k,n)=(4,8)$ case of interest for amplitudes. The facets of $\mathcal{C}(4,8)$ have some normal rays that point in non-cluster directions, and it is natural to ask for the generating function for the canonical basis elements associated to all points along each such ray. This will motivate an association between the poles of this generating function and symbol letters, revealing a remarkable connection between $\mathcal{C}(4,8)$, the canonical basis, and the quadratic equation associated with the four-mass box symbol letters.

\subsection{A Rank-2 Example}

Before jumping into the intricacies of $\Gr(4,8)$ let us warm up by illustrating many of the salient points with the simplest example of an infinite-type cluster algebra. Largely following~\cite{sherman2004positivity}, consider the rank-2 cluster algebra with initial exchange matrix
\begin{align}
B = \begin{pmatrix} 0 & -2 \\ 2 & 0 \end{pmatrix}
\end{align}
whose cluster variables $x_n$ are determined in terms of those of the initial cluster $(x_1, x_2)$ by
\begin{align}
\label{eq:mutation}
x_{n+1} = \frac{1 + x_n^2}{x_{n-1}}\,.
\end{align}
For $a,b \in \mathbb{Z}$, the ${\bf g}$-vector of a monomial $x_1^a x_2^b$ in the initial variables is defined to be $(a,b)$. As mentioned above, for general cluster variables, the {\bf g}-vector is given by the exponent vector in the Laurent expansion which is smallest by the partial ordering specified by $B$. For example, two cluster variables are
\begin{align}
x_4 =
\frac{1}{x_2} +
\frac{1}{x_1^2x_2}
+ 2 \frac{x_2}{x_1^2} +
\frac{x_2^3}{x_1^2}
\end{align}
and
\begin{align}
x_0 = \frac{x_1^2}{x_2} + \frac{1}{x_2}\,,
\end{align}
where in each case the terms are written in increasing order with respect to $\preceq$. We see that the ${\bf g}$-vector of $x_4$ is $(0,-1)$. In general, it is easy to work out that the {\bf g}-vector associated to $x_n$ is $(n{-}4,3{-}n)$ for $n \geq 3$ and $(2{-}n,n{-}1)$ for $n \leq 2$. The union of ${\bf g}$-vector cones \emph{almost} covers all of $\mathbb{R}^2$, but they accumulate along a single missing ray generated by $(1,-1)$. It is thus clear that cluster monomials don't provide a basis for the full cluster algebra. For instance, the smallest monomial in the product $x_1 x_4$ has ${\bf g}$-vector equal to $(1,-1)$, and so can't be written as a sum of cluster monomials. What is needed to complete a basis for the cluster algebra?

It is easy to give an elementary answer to this question in this very simple example. We begin by noting that the mutation relation~(\ref{eq:mutation}) may be recast into the form of a recurrence relation
\begin{align}
x_{n+1} = A x_n - x_{n-1} ~
\mbox{with} ~
A = \frac{x_1}{x_2} + \frac{1}{x_1 x_2} + \frac{x_2}{x_1}\,.
\label{eq:chebyshev}
\end{align}
In fact $x_1,x_2$ can be replaced by any $x_n,x_{n+1}$ in the expression for $A$; it is an invariant across all clusters. We can use this generating function to give an explicit expression for all cluster variables. If we define the generating function $X(t) = \sum_{k \ge 0} x_k t^k$, then the recurrence relation implies that $(1 - A t + t^2) X(t) = x_0 + t (x_1 - A x_0)$, and hence
\begin{align}
X(t) = \frac{x_0(1 - A t) + x_1 t}{1 - A t + t^2}\,.
\end{align}
By factoring the quadratic equation in terms of its roots $R_{\pm}$, we can also express $x_n$ as
\begin{align}
x_n = \frac{x_0 (R_-^{n+1} - R_+^{n+1}) + (x_1 - A x_0)(R_-^n - R_+^n)}{R_- - R_+}
\end{align}
where
\begin{align}
R_\pm = \frac{A \pm \sqrt{A^2 - 4}}{2}\,.
\end{align}
Here we see the first occurrence of an interesting quad\-rat\-ic equation and its associated roots in the generating function $X(t)$ for the cluster variables themselves.

The variable $A$ has a deeper significance. Note that while e.g.~$x_1 x_4$ can't be expressed as a sum of cluster monomials, we can write $x_1 x_4 = A + x_2 x_3$, suggesting that $A$ should be considered an element of the basis. Its {\bf g}-vector is $(1,-1)$, which lies along the missing ray, so it is a basis element that is not a cluster variable.

What basis elements should we associate with the other integer points $(p,-p)$ along this ray? Most naively, in analogy with the case of cluster variables, we might think that these should just be the powers $A^p$. But the variable $A$ has a very interesting and peculiar property that suggests this is the wrong answer. Cluster variables are ``critically positive", in the sense that they approach zero on some boundaries of the positive part of the cluster variety. But this is not true of $A$! Note that $A = r + 1/r + 1/(x_1 x_2)$ where $r=x_1/x_2$, and thus $A \geq 2$. $A$ is thus ``overpositive", and can't reach zero on any boundary of the positive part. Relatedly, we can't understand the positivity of $A - 2$ in the way that one familiarly understands positivity of polynomials in cluster variables. Usually, an expression can be determined to be positive simply by expanding in the cluster variables of an initial seed and seeing that all terms in the Laurent expansion are positive. But that is not the case for $A-2$; this expression is positive despite the appearance of a negative sign in its Laurent expansion simply because $A^2 - 4 > (r+1/r)^2 - 4 = (r - 1/r)^2 > 0$.

Motivated by this observation, we say that an element $x$ of the algebra is \emph{positive} if it is positive-valued when $x_1, x_2 > 0$ and that a positive $x$ is \emph{Laurent positive} (with respect to the initial cluster) if it is a linear combination of initial cluster monomials with positive coefficients; otherwise $x$ is \emph{nontrivially positive}. Finally we say that $x$ is \emph{overpositive} if there exists a positive $y$ such that $x/y$ is non-constant and $\min(x/y)>0$. Note that
\begin{align}
A^2 = \frac{2}{x_1^2} + \frac{2}{x_2^2}
+ \frac{1}{x_1^2 x_2^2} + \frac{x_1^2}{x_2^2}
+ \frac{x_2^2}{x_1} + 2
\end{align}
so we see that $A^2 - z$ is Laurent positive only for $z \le 2$ (this is another way to see that $A$ is overpositive). It is still positive for $2 < z \le 4$ since we have already seen that $A \ge 2$, and an important consequence of this nontrivial positivity is that $R_\pm$ are both real and positive-valued in the positive part of the cluster varitey. Finally, it is easy to see that $A^2 - z$ is not positive for $z > 4$.

Moving to higher powers, it is natural to ask for some basis $T_p(A)$ of polynomials in $A$, beginning with $T_0(A) = 1$, $T_1(A) = A$ and $T_2(A) = A^2 - 2$, such that $T_p(A)$ is maximally Laurent positive (that means, with no further subtractions possible). Using $A = r + 1/r + 1/(x_1 x_2)$, such polynomials can be determined by requiring $G_p(z + 1/z) = z^p + 1/z^p$. The generating function for basis elements of this form along the missing ray, $g(t) = \sum_{p \ge 0} G_p(A) t^p$, is given by $g(t) = (1 - t^2)/(1 - A t + t^2)$. Note the second appearance of the the same quadratic polynomial in $A$ we saw earlier, this time not in computing cluster variables, but more fundamentally as generating the basis elements for this ``critically Laurent positive" basis along the missing $(1,-1)$ ray.

It is also natural to consider another basis $F_p$ with generating function $f(t) = (1 - A t + t^2)^{-1}= \sum_{p \ge 0} F_p(A) t^p$, which is expected to be the analogue of Lusztig's canonical basis for this rank 2 cluster algebra. We will not attempt to explain the deep significance of this basis here; for now, we simply wish to emphasize that the nontriviality of the generating function for both the bases we have highlighted is associated with the surprising overpositivity of the non-cluster variable $A$ associated with the first integer point on the non-cluster ray. We also emphasize that while the generating functions $f(t)$ and $g(t)$ clearly differ, they have the same poles in $t$; we expect this will be true of any suitably reasonable basis.

\subsection{A Definition of Cluster Algebraic Functions}

Let ${\mathcal A}$ be a cluster algebra of rank $d$, let ${\bf X} \simeq \mathbb{Z}^d$ be a lattice that parameterizes bases of ${\mathcal A}$, and let $\mathcal{B}({\bf g})$ be a basis element associated to the lattice point ${\bf g} \in {\bf X}$. For a ray in ${\bf X}$ with integer points $1 \cdot {\bf g}, 2 \cdot {\bf g}, \ldots$ (note that we always take ${\bf g}$ to be the first integer point along its ray) we define the generating function
\begin{align}
\label{eq:genfuncdef}
f_{\bf g}(t) = \sum_{k \ge 0} \mathcal{B}(k {\bf g}) t^k\,.
\end{align}
In general $f_{\bf g}(t)$ depends on the choice of basis, but if ${\bf g}$ lies in the cluster fan of ${\mathcal A}$ then $\mathcal{B}(k {\bf g}) = \mathcal{B}({\bf g})^k$, where $\mathcal{B}({\bf g})$ is the cluster monomial associated to ${\bf g}$, and hence
\begin{align}
\label{eq:simplepole}
f_{\bf g}(t) = \frac{1}{1 - t \mathcal{B}({\bf g})}\,.
\end{align}

Motivated by~(\ref{eq:simplepole}), in cases where $f_{\bf g}(t)$ is a rational function of $t$ we denote the roots of $1/f_{\bf g}(1/t)$ by $R({\bf g})$. We conjecture that these are always positive on ${\mathcal A}_{> 0}$, the positive part of ${\mathcal A}$. We call $R({\bf g})$ the \emph{cluster algebraic function} associated to the ray generated by ${\bf g}$. If ${\bf g} \in {\bf X}$ is a ${\bf g}$-vector of the cluster algebra ${\mathcal A}$ then $R({\bf g})$ contains a unique element, the cluster variable associated to ${\bf g}$, but if ${\bf g}$ lies outside the cluster fan of ${\mathcal A}$ then $R({\bf g})$ is a finite collection of algebraic functions of cluster variables. Further aspects of such functions will be discussed in~\cite{toappear}.

\section{IV. $\Gr(4,8)$ and the Four-Mass Box}

Here we apply the proposal just introduced to the polytopes constructed in section II. As reported there, we find that the normal rays to 356 facets of $\mathcal{C}(4,8)$ are generated by ${\bf g}$-vectors of the $\Gr(4,8)$ cluster algebra, and therefore are naturally associated to 356 of its cluster variables. The other 4 facets have normal rays generated by
\begin{align}
\begin{split}
{\bf g}_1 &= (-1, 1, 0, 1, 0, -1, 0, -1, 1)\,, \\
{\bf g}_2 &= (0, -1, 0, -1, 0, 1, 0, 1, 0)\,, \\
{\bf g}_3 &= (-1, -1, 1, -1, 2, 0, 1, 0, -1)\,, \\
{\bf g}_4 &= (1, 0, -1, 0, -2, 1, -1, 1, 1)\,.
\end{split}
\label{eq:fourmissing}
\end{align}
The first two of these are also normal rays to $\mathcal{C}^\dagger(4,8)$. We adopt Corollary 7.3 of~\cite{CDFL} as a (conjectural) way to assign a canonical basis element to any lattice point in $\mathbb{R}^d$, regardless of whether it is in the cluster fan. In the notation defined in that paper, the semistandard Young tableaux $T_1, \ldots, T_4$ associated to~(\ref{eq:fourmissing}) are respectively
\begin{align}
\begin{ytableau}
1 & 3 \\
2 & 5 \\
4 & 7 \\
6 & 8
\end{ytableau}\,,
\qquad
\begin{ytableau}
1 & 2 \\
3 & 4 \\
5 & 6 \\
7 & 8
\end{ytableau}\,,
\qquad
\begin{ytableau}
1 & 1 & 2 & 4 \\
2 & 3 & 3 & 6 \\
4 & 5 & 5 & 7 \\
6 & 7 & 8 & 8
\end{ytableau}\,,
\qquad
\begin{ytableau}
1 & 1 & 2 & 3 \\
2 & 4 & 4 & 5 \\
3 & 6 & 6 & 7 \\
5 & 7 & 8 & 8
\end{ytableau}\,.
\end{align}
Let us first consider $T_1$, which is the same as the $T_4$ considered in Example 8.1 of~\cite{CDFL}, where the variables associated to the first two points along the ray generated by ${\bf g}_1$ were computed as
\begin{align}
\begin{split}
\mathcal{B}(1 \cdot {\bf g}_1) &= \ch(T_1) = A\,,\\
\mathcal{B}(2 \cdot {\bf g}_1) &= \ch(T_1 \cup T_1) = A^2 - B
\label{eq:firsttwo}
\end{split}
\end{align}
in terms of the quantities
\begin{align}
\begin{split}
A &= \langle 1256 \rangle \langle 3478\rangle - \langle 1278 \rangle \langle 3456 \rangle - \langle 1234 \rangle \langle 5678 \rangle\,, \\
B &= \langle 1234 \rangle \langle 3456 \rangle \langle 5678 \rangle \langle 1278 \rangle\,.
\end{split}
\label{eq:abdef}
\end{align}
We have computed the next variable along this ray,
\begin{align}
\mathcal{B}(3 \cdot {\bf g}_1) = \ch(T_1 \cup T_1 \cup T_1) = A^3 - 2 A B\,.
\label{eq:third}
\end{align}
Based on~(\ref{eq:firsttwo}) and~(\ref{eq:third}) we conjecture that
\begin{align}
\label{eq:genfuncconjecture}
f_{{\bf g}_1}(t) = \sum_{k \ge 0} \ch(T_1^{\cup k}) t^k = \frac{1}{1 - t A + t^2 B}\,.
\end{align}
Note that this encapsulates an infinite number of predictions about the behavior of the canonical basis along the direction ${\bf g}_1$. According to the proposal outlined in the previous section, the variables associated to this ray are therefore
\begin{align}
R({\bf g}_1) = \frac{A \pm \sqrt{A^2 - 4 B}}{2}\,.
\end{align}
Exactly as in the rank two toy example we studied in the previous section, using the representation of $A, B$ in terms of initial cluster variables it is straightforward to check that $A$ is overpositive and $\Delta \equiv A^2 - 4 B$ is positive, even though $A^2 - r B$ is Laurent positive only for $r \le 2$ and not positive if $r > 4$. This can be seen by noting that expanding in terms of initial cluster variables, we have $A = x + y + \cdots$, where $x,y$ are monomials such that $B = x y$. This shows that $A^2 - 4B$ is positive and that $A^2 - 2 B$ is Laurent positive. Furthermore, it can be checked that the exponent vectors of $x,y$ are separated from the exponent vectors of all the other monomials in $A$ by a hyperplane. This means that we can scale the cluster variables in such a way that $x,y$ dominate by arbitrarily large factors relative to the other monomials in $A$, and shows that we can make $A^2 - r B$ negative for $r>4$, so $r=4$ is the critical case.

Even more exciting is the fact that $\sqrt{\Delta}$ is precisely one of the two cyclic incarnations of the square root that appears in the four-mass box integral~\cite{Hodges1977,tHooft:1978jhc}; its cyclic partner comes from the ray generated by ${\bf g}_2$. These are known to be the only square roots that appear in the symbol alphabets of the one-loop N${}^2$MHV~\cite{Britto:2004nc} and two-loop NMHV amplitudes~\cite{Zhang:2019vnm}; they are also expected to appear in MHV amplitudes at three loops and beyond~\cite{Prlina:2017tvx}.

Finally we turn to ${\bf g}_3$ and its cyclic partner ${\bf g}_4$. It is straightforward to compute the variable $G = \mathcal{B}(1 \cdot {\bf g}_3) = \ch(T_3)$ associated to the lattice point ${\bf g}_3$. Remarkably we find that $G$, which has torus weight 2 in each of the eight $Z_i$, is related to $A$ by a braid element~\cite{FraserBraid} of the $\Gr(4,8)$ cluster modular group. Under the same braid transformation we find that $B \mapsto B'$ where
\begin{align}
B' = B \,\langle 2345 \rangle \langle 4567 \rangle \langle 1678 \rangle
\langle 1238 \rangle\,.
\end{align}
Therefore $\Delta$ maps to $\Delta' = G^2 - 4 B'$, which again is nontrivially positive. While the presence of the additional root $\sqrt{\Delta'}$ (and its cyclic partner) associated to $\mathcal{C}(4,8)$ has a simple and beautiful mathematical origin, it is unclear whether these roots play any role in physics; for example, whether they correspond to the Landau singularities. This may be another sign (beyond the consideration of parity discussed above) that $\mathcal{C}^\dagger(4,8)$ (which lacks these extra roots) is more relevant to the physics of MHV amplitudes than $\mathcal{C}(4,8)$.

\section{Outlook}

Our work raises a number of related questions. On the mathematical side, it would be interesting to verify the conjecture~(\ref{eq:genfuncconjecture}) and to explore the corresponding generating functions at higher $n$ and $k$. More generally, it would be interesting to understand under what circumstances the generating function defined in~(\ref{eq:genfuncdef}) is rational (for example, is this true in $\Gr_+(k,n)$ for all ${\bf g}$?) and to prove our conjecture that in such cases the roots of its denominator are always positive.

On the physics side it would be interesting to determine if the 272 cluster variables associated to $\mathcal{C}^\dagger(4,8)$, together with the algebraic letters of four-mass-box type, indeed constitute the all-loop symbol alphabet of the 8-particle MHV amplitude. Evidence in support of this suggestion could be provided either by explicit computation or by using them as a bootstrap ansatz. Also, we have so far only discussed the crudest relation between polytopes and cluster algebras, according to which symbol letters of the latter are related to facets of the former. For finite algebras this connection was first observed in~\cite{Golden:2013xva} but in recent years much finer connections have been explored, for example the observation~\cite{Drummond:2017ssj} that two cluster variables can appear next to each other in a symbol only if the corresponding facets intersect. From~(\ref{eq:unnamed}) we see that $\mathcal{C}^\dagger(4,7)$ has 385 codimension-2 faces, suggesting that only 385 distinct pairs of adjacent symbol entries (all of the ones tabulated in~\cite{Drummond:2017ssj} except for those described in equation (12b)) appear in 7-particle MHV amplitudes. This is consistent with all evidence available to date~\cite{Southampton}. It would be interesting to explore this type of finer structure for $n>7$, and especially to understand the ``cluster adjacency'' properties of algebraic symbol letters.

We leave the most ambitious question for last. Supposing that an appropriate non-perturbative geometry for pSYM theory is indeed provided by a construction
of the type that we have described, what is the non-perturbative question we should ask of this space, whose answer gives a scattering amplitude?

\section{Acknowledgements}

It is a pleasure to thank J.~Drummond, J.~Foster, {\" O}.~G{\"u}rdo{\u g}an, S.~He, C.~Kalousios and A.~Volovich for stimulating conversations, and C.~Fraser and J.-R.~Li for helpful correspondence. This work was supported in part by DOE grants DE-SC0009988 (NAH) and DE-SC0010010 Task A (MS).

\end{document}